\documentclass[%
longbibliography,
nofootinbib,
nobibnotes,
bibnotes,
 amsmath,amssymb,
aps,
pra,
twocolumn,
]{revtex4-2}

\usepackage{graphicx}
\usepackage{dcolumn}
\usepackage{bm}
\usepackage{nccmath,amsmath}
\usepackage{amsfonts}
\usepackage{mathtools}
\usepackage{braket}
\usepackage{bbm}
\usepackage{ulem}

\usepackage{xcolor}
\usepackage{txfonts}

\usepackage[font=small,labelfont=bf,
   justification=Justified,
   format=plain]{caption}
\usepackage{subcaption}

\begin{document}

\preprint{APS/123-QED}

\title{Experimental characterization of coherent and non-Markovian errors\\using tangent space decomposition}%



\author{Elia Perego$^{1^{*,\dagger}}$}
\author{Andrea Rodriguez-Blanco$^{2^{*}}$ }
\author{K. Birgitta Whaley$^{2}$ }
\author{Bharath Hebbe Madhusudhana$^{3, 4}$ }

\def\thefootnote{*}\footnotetext{These authors contributed equally to this work.}
\def\thefootnote{$\dagger$}\footnotetext{Current address: Eniquantic S.p.A., via Ostiense 72, 00154 Roma, Italy}

\affiliation{$^{1}$ Department of Physics, University of California, Berkeley, California 94720, USA}
\affiliation{$^{2}$ Department of Chemistry, University of California, Berkeley, California 94720, USA }
\affiliation{$^{3}$ New Mexico Consortium, Los Alamos, NM-87544, United States}
\affiliation{$^{4}$ MPA-Quantum, Los Alamos National Laboratory, Los Alamos, NM-87544, United States}
\date{\today}

\begin{abstract}
Accurate characterization of coherent and non-Markovian errors remains a central challenge in quantum information processing, as conventional benchmarking techniques typically rely on Markovian and time-independent noise assumptions. In practice, however, quantum devices exhibit both systematic coherent miscalibrations and temporally correlated fluctuations, which complicate error diagnosis and mitigation. Here, we apply a technique based on tangent-space decomposition to characterize such error in single-qubit quantum gates implemented on a trapped ion platform. Small imperfections in a quantum operation are treated as perturbations of the target quantum map, represented as tangent vectors in the space of quantum channels. This formulations enables a natural decomposition of the deviation into three components corresponding to coherent, Markovian and non-Markovian processes.The relative weights of these components provide a quantitative measure of the contribution from each type of error mechanism, directly from a single tomographic snapshot. We experimentally validate this method on a single-qubit gates implemented on a trapped $^{40}$Ca$^+$ ion, where control is achieved through laser-driven optical transitions. By analyzing experimentally reconstructed process matrices, expressed in the Pauli Transfer Matrix and Choi representations, we identify and quantify non-Markovian effects arising from controlled injection of slow fluctuations in the experimental environment. We also characterize deterministic coherent miscalibrations using the same technique. This approach provides a physically transparent and experimentally accessible tool for diagnosing complex error sources in quantum control systems.
\end{abstract}

\keywords{Coherent errors, non-Markovian errors, Trapped ions} 
\maketitle




\section{Introduction}
The characterization and stabilization of open quantum systems plays an important role in the development of fault-tolerant (FT) quantum information processing technologies.
\begin{figure}
    \centering
    \includegraphics[width=\linewidth]{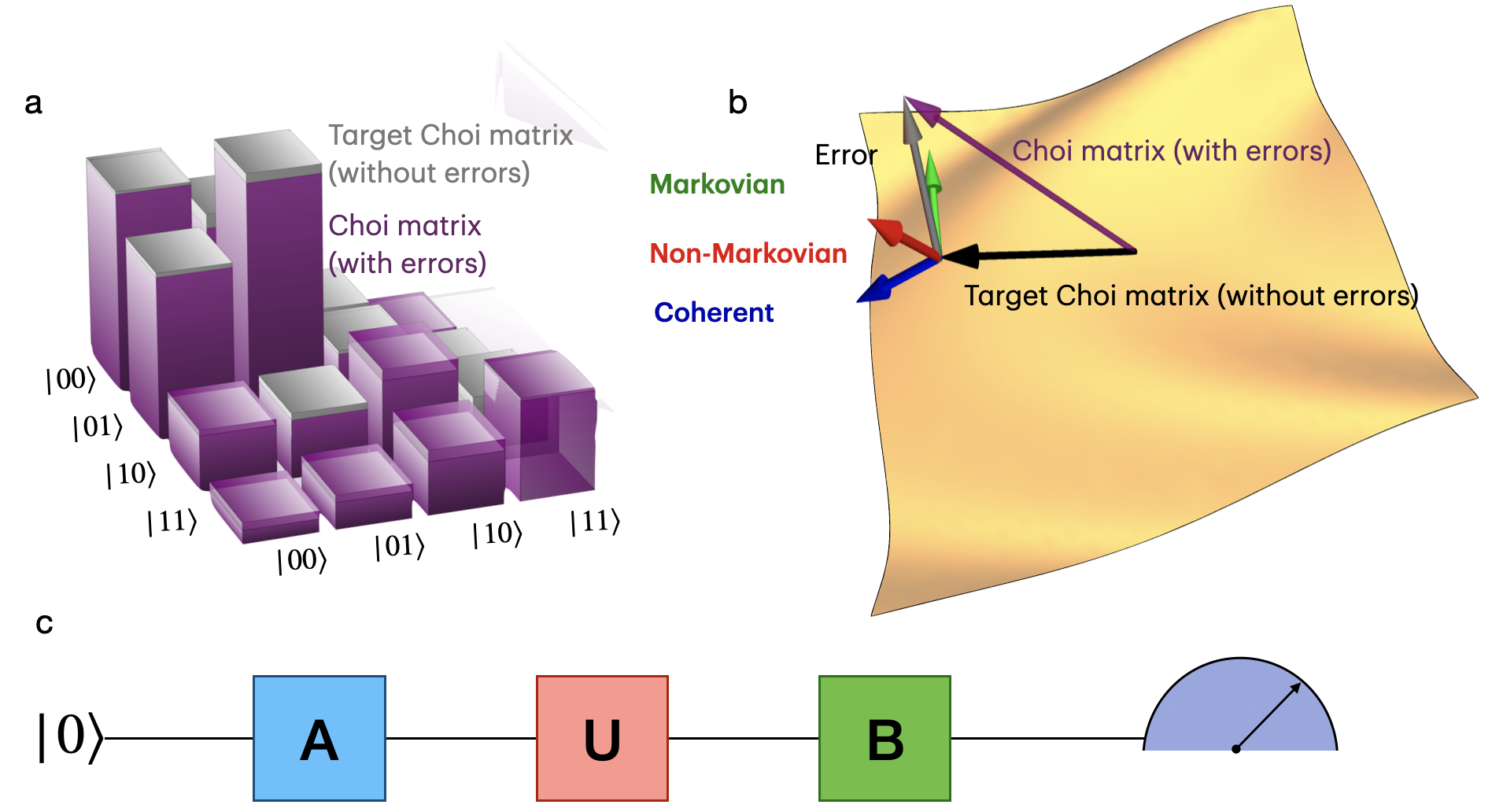}
    \caption{\textbf{Tangent space decomposition:} \textbf{a.} The Choi matrix is a $4\times 4$, Hermitian, positive semi-definite matrix that characterized the quantum process completely. It can be measured using a process tomography. \textbf{b.} The space of Choi matrices form a manifold and the errors in quantum control are a perturbation on the surface of the manifold. The difference between the target Choi matrix and the measured Choi matrix can be decomposed into three components corresponding to three distinct sources of error. \textbf{c.} Experimental sequence for process tomography of a target unitary gate $U$ on a single qubit. The gates $A$ and $B$ are varied across $12$ combinations (see text) to recover the Choi matrix completely.
    }
    \label{Fig1}
\end{figure}
Markovian error models are usually employed in the development of theoretical and experimental approaches to error mitigation and fault-tolerant quantum computing, 
motivated by the fact that fault-tolerant thresholds are typically derived assuming stochastic, memoryless Markovian errors \cite{doi:10.1137/S0097539799359385,10.5555/2011665.2011666,Nielsen_Chuang_2010,PhysRevA.77.012307,doi:10.1126/science.1145699}.
However, in practice, quantum processors are affected by factors such as temporally correlated noise and memory effects that give rise to non-Markovian dynamics, as well as by systematic coherent errors during quantum logic operations. Neither of these are fully captured by conventional benchmarking techniques which typically rely on Markovian and time-independent noise assumptions \cite{PhysRevA.77.012307,doi:10.1126/science.1145699, PhysRevA.87.062119, Nielsen2021gatesettomography}. These beyond-Markovian effects jointly complicate both accurate error characterization and the validation of error-correction protocols. 

Coherent errors arising from systematic miscalibrations during quantum logic operations constitute an important class of control errors. Such errors are typically inferred from calibration routines or from deviations in reported gate fidelities. However, widely used benchmarking metrics such as the average gate fidelity are only second-order sensitive to small coherent miscalibrations \cite{PhysRevLett.117.170502}, and randomized benchmarking protocols tend to average away coherent structure \cite{PhysRevLett.119.130502}. Several works have focused on developing methods to better characterize or mitigate coherent errors,  including noise tailoring via randomized compiling \cite{PhysRevA.94.052325}, cycle benchmarking \cite{Erhard2019}, and efficient noise learning techniques that explicitly identify unitary error components \cite{Harper2020}. However, despite these advances, experimentally efficient methods that directly separate coherent, Markovian and non-Markovian contributions at the gate level remain limited.

Early attempts to detect and quantify non-Markovianity followed two main criteria: (i) the temporary information backflow from the environment to the system increases the trace distance between states, quantified by the Breuer-Laine-Piilo (BLP) measure \cite{
PhysRevLett.103.210401}, and (ii) the failure of completely positive (CP)-divisibility of a dynamical map, quantified by the Rivas-Huelga-Plenio (RHP) measure \cite{PhysRevLett.105.050403}. A complementary ``single-snapshot'' method \cite{PhysRevLett.101.150402} presents necessary and sufficient conditions to decide whether the quantum channel is Markovian or non-Markovian within the single snapshot. However, most of the current benchmarking methods use multiple time characterization protocols.

In recent years, there has been a remarkable focus on enhancing the efficiency of benchmarking and characterization protocols, together with the definition of appropriate quality measures. Multi-time frameworks reconstruct memory structure explicitly. Thus, the process-tensor formalism \cite{PhysRevA.97.012127}, which originated as a generalization from the Feynman-Vernon influence functional \cite{FEYNMAN1963118, 10.21468/SciPostPhys.18.1.024}, provides a complete multi-time description of open quantum dynamics by encoding all temporal correlations between system and environment. Its experimental reconstruction through process-tensor tomography \cite{PRXQuantum.3.020344}, and its discrete counterpart, the transfer tensor method \cite{PhysRevLett.112.110401}, enable the identification and quantification of non-Markovianity by revealing memory effects and information backflow. 
Parallel to this approach, noise spectroscopy methods connect observables to bath spectra. Filter-functions \cite{PhysRevLett.87.270405,PhysRevLett.98.100504, PhysRevLett.107.230501, Bylander2011, PhysRevLett.109.020501} constitute spectroscopic diagnostic tools that characterize how known control sequences filter external noise in the frequency domain. However, while the filter-function formalism provides an experimentally efficient way to reconstruct the spectral characteristics of environmental noise and thus infer classical non-Markovian correlations, it can not fully reconstruct or quantify quantum memory effects that multi-time frameworks do. Recent frame-based filter-function \cite{PhysRevLett.116.150503, PRXQuantum.2.030315} approaches extend the standard formalism to enable study of colored and time-varying (non-stationary) noise, allowing the analysis of non-Markovian effects and offering a practical way to capture temporally correlated and cross-axis error processes in experiments. Varona et al. \cite{Varona2025} have introduced a different approach with Lindblad-like quantum tomography, which extends conventional Lindblad tomography to non-Markovian quantum dynamical maps, enabling the reconstruction of effective time-local generators even in the presence of temporally correlated noise. Furthermore, while Lindblad-like quantum tomography provides a time-local, continuous description of non-Markovian dynamics by reconstructing effective generators that capture temporally correlated noise, similar ideas can be applied at the discrete, gate-based level of quantum circuits. Thus, context-aware Gate Set Tomography (GST) \cite{viñas2025contextawaregatesettomography} analogously extends conventional GST \cite{PhysRevA.87.062119,greenbaum2015introductionquantumgateset,Nielsen2021gatesettomography} by incorporating history dependence into the model, thereby accounting for gate-level non-Markovian effects that arise from correlations between successive operations.
 
In this work, we present a novel geometry-based framework for characterizing errors in single-qubit operations using a tangent space decomposition on the manifold of completely positive trace-preserving (CPTP) maps.
Small deviations between the implemented and ideal operations are considered as tangent vectors that can be uniquely decomposed in three physically interpretable components: coherent, Markovian and non-Markovian. In principle, this approach allows the relative contribution of each error process to be quantified directly from a single tomographic snapshot, without assuming any particular noise model, memory kernel, or modality of temporal dynamics. 
An important feature of the tangent space  method is that its geometry-based error decomposition technique  can be generalized to any number of qubits without incurring a significant increase in the number of measurement settings or samples. This approach
therefore fills a methodological gap between general multi-time or spectral noise frameworks and the experimentally challenging and expensive gate-level (or generator-level) tomographies, providing a low-overhead diagnostic tool for identifying and quantifying coherent errors and non-Markovian noise in real quantum devices. In this paper we make a first experimental application of the theory, to single-qubit gates on a trapped ion quantum computing platform.

The remainder of the paper is organized as follows. In Sec.~\ref{Sec:tangent}, we introduce the theoretical framework of tangent-space decomposition for a single qubit, and define the corresponding coherent, Markovian, and non-Markovian subspaces. In Sec.~\ref{Sec:experiment}, we introduce the experimental implementation of the tomography protocol on a single trapped-ion qubit, and experimentally demonstrate how the method enables the quantitative identification of both systematic and temporally correlated errors.  Sec IV summarizes and outlines possible extensions of this approach to multi-qubit systems and higher-dimensional Hilbert spaces.


\section{Tangent space decomposition}\label{Sec:tangent}
In this section, we develop the theory of tangent space decomposition for a single-qubit system and describe how it can be applied to characterize quantum control errors. 
The goal is to identify and separate different physical sources of error --- coherent miscalibrations, stochastic Markovian noise, and temporally correlated non-Markovian effects --- directly from experimentally reconstructed quantum processes. 
To achieve this, we adopt a geometric description of quantum channels that allow small experimental imperfections to be treated perturbatively.
A CPTP map $\Phi$ acting on a single qubit, i.e. the space of $2\times 2$ density matrices, can be represented as a $4\times 4$ complex matrix in several ways, depending on the basis chosen for the space of density matrices. A useful representation is the Choi matrix (CM) \cite{CHOI1975285,Nielsen_Chuang_2010}.
The Choi matrix has rich information-theoretic insights, such as the fact that its partial traces encode the information hierarchy of the map. It also highlights a close parallel between quantum states and quantum operations. Most importantly here, it simplifies the notion of complete positivity of $\Phi$, to the requirement that the Choi matrix be positive semidefinite (PSD).
The Choi representation is also particularly suited for establishing a geometric framework: physical quantum channels correspond to PSD matrices subject to linear trace constraints, forming a smooth manifold embedded in a linear vector space. This structure enables a natural perturbative treatment of small deviations between ideal and experimentally implemented operations.

The CM corresponding to a map $\Phi$ is defined as
\begin{equation}\label{eq:CM}
    \rho^{\Phi} = \sum \Phi(\ket{i}\bra{j})\otimes \ket{i}\bra{j}
\end{equation}
Here, $\ket{i}$ are the basis elements of $\mathbb{C}^2$ and $\ket{i}\bra{j}$ form a basis for $2\times 2$ complex matrices, including density matrices. \\
A completely positive (CP) map $\Phi$ corresponds to a Hermitian, PSD matrix $\rho^{\Phi}$. The set of positive semidefinite Choi matrices forms a convex set in the real
vector space of Hermitian matrices~\cite{Watrous2018, convex_cone}. If $\Phi$ is also trace-preserving, one of the partial traces of $\rho^{\Phi}$ is equal to $2\times 2$ identity: $\text{Tr}_1[\rho^{\Phi}]=\mathbbm{1}$. 
While the space of CP maps is a $16$ dimensional real manifold, trace preserving maps form a $12$ dimensional submanifold, $\mathcal M$. This subspace is a linear slice of the set of PSD matrices - namely, the intersection of the PSD set with the linear space defined by $\text{Tr}_1[\rho^{\Phi}]=\mathbbm{1}$ \cite{Watrous2018}. The geometric picture is central to this approach because it allows to interpret experimental imperfections as small displacements on the manifold of CPTP maps.

Given an experimental implementation of a CP map, 
we consider two CMs: $\rho^{\Phi}$ and $\rho^{\Phi}_{\text{expt}}$. The former is the exact CM corresponding to $\Phi$ (Eq.~\ref{eq:CM}) and the latter is the map that is implemented experimentally, which is often measured using a tomographic protocol (see Sec.~\ref{subsec:rcm}). We represent the pair $(\rho^{\Phi}, \rho^{\Phi}_{\text{expt.}})$ as $(\rho^{\Phi}, \Delta \rho^{\Phi})$ where $\Delta \rho^{\Phi}= \rho^{\Phi}_{\text{expt.}}-\rho^{\Phi}$ is the error. When the error $ \Delta \rho^{\Phi}$ is sufficiently small, it can be represented as a \textit{tangent vector} to the manifold $\mathcal M$ at the point $\rho^{\Phi}$, such that $ \Delta \rho^{\Phi} \in \mathcal T_{\rho^{\Phi}} \mathcal M$, where $T_{\rho^{\Phi}} \mathcal M$ denotes the $12$-dimensional tangent space on $\mathcal M$. The tangent space description provides a local, linear approximation to the space of physical quantum channels, enabling a systematic classification of error mechanisms according to their first-order effects.

We will show that this $12$-dimensional tangent space can be written as a direct sum of three subspaces:
\begin{equation}\label{eq:TSD}
    T_{\rho^{\Phi}} \mathcal M =   V_{\text{Mark.}}\oplus \mathcal V_{\text{NonMark.}}\oplus\mathcal V_{\text{Cohr.}}.
\end{equation}
Here, $\mathcal V_{\text{Cohr.}}$ is a $3-$dimensional subspace corresponding to coherent errors, i.e., errors arising from systematic under- or over-rotations. This space is spanned by rotations about $x, y$ and $z$, axes generated by the three Pauli matrices. The subspace  $\mathcal V_{\text{Mark.}}$ is also $3-$dimensional and corresponds to Markovian (memoryless) noise processes. The remaining $6-$dimensional subspace $\mathcal V_{\text{NonMark.}}$ captures all other error contributions, which we associate with non-Markovian effects. 
Correspondingly, any small error $\Delta \rho^{\Phi}$ can be written as a sum of three terms (see Fig.~\ref{Fig1}):
\begin{equation}
    \Delta \rho^{\Phi} = \Delta \rho^{\Phi}_{\text{Mark.}}+\Delta \rho^{\Phi}_{\text{NonMark.}}+\Delta \rho^{\Phi}_{\text{Cohr.}}
\end{equation}

This decomposition is based on common physical processes and a clear mathematical distinction between the error mechanisms.

\subsection{From Choi to PTM for error characterization}
While the Choi matrix provides a natural geometric framework for defining physical quantum channels and their tangent space, it is not the most convenient representation to identify the physical origin of different error mechanisms. In particular, the distinction between coherent errors, stochastic Markovian, and non-Markovian effects becomes most transparent in the Pauli transfer matrix (PTM) representation, which describes the action of a channel directly on the Bloch vector. The Pauli transfer matrix and Choi representation are related by a linear change of basis (see Appendix \ref{app:CMtoPTM}). For a single-qubit channel $\Phi$, the Choi matrix can be expressed in the Pauli basis as: 

\begin{equation}\label{eq:Choi_PTM}
\rho^{\Phi}=\frac{1}{2}\sum_{\alpha , \beta}T_{\alpha, \beta}\sigma_{\alpha}\otimes\sigma_{\beta}^{T},
\end{equation}

while the inverse relation is given by 

\begin{equation}\label{eq:PTM_Choi}
T_{\alpha,\beta}={\rm Tr}[\rho^{\Phi}(\sigma_{\alpha}\otimes\sigma_{\beta}^T )].
\end{equation}

This equivalence allows us to conveniently formulate the measurement model and error decomposition in the PTM representation, while enforcing complete positivity in the Choi matrix. We now switch to the PTM picture to classify and quantify the different components of the tangent-space error decomposition introduced above. 

\subsubsection{Markovian errors}
We begin the classification of errors by identifying Markovian contributions, corresponding to time-independent, memoryless noise processes. These errors are most naturally described in the Pauli transfer matrix representation \cite{PhysRevLett.109.060501, PhysRevA.87.062119,greenbaum2015introductionquantumgateset}, which directly encodes how the expectation values of the Pauli operators—and hence populations and coherences—are transformed by the channel.
In the PTM representation, a single-qubit quantum channel is described by a real 4x4 matrix acting on the Pauli vector of the input state. For a trace-preserving map, the PTM takes the block form 

\begin{equation}\label{eq:PTM_matrix}
T=\begin{pmatrix}
1 & \vec{0} \\
\vec{t} & R 
\end{pmatrix},
\end{equation}

where the upper-left element enforces trace preservation, the vector  $ \vec{t}$ captures  shifts of the Bloch vector, and the $3\times 3$ matrix $R$ describes the linear transformation of the Bloch vector. For an ideal unitary operation $\vec{t}=\vec{0}$ and $R\in SO(3)$, corresponding to a pure rotation of the Bloch sphere. Deviations from this decomposition encode different physical error mechanisms. 

We denote by $T_{\rm ideal}$ and $T_{\rm expt.}$ the PTMs corresponding to the ideal and experimentally implemented maps, respectively. To quantify the deviation between these, we introduce a distance measure on the space of PTMs. Throughout this work, we employ the Frobenius norm $||T||_{F}^2={\rm Tr}(TT^T)$ as a Euclidean metric on the space of Pauli transfer matrices, which are real matrices and thus admit a natural quadratic distance measure. The squared Frobenius norm is the sum of the squares of all individual elements in the matrix, and for the PTM can be written as

\begin{equation}
\begin{split}
\epsilon^2&=||T_{\rm ideal}-T_{\rm expt}||^2 \\
&=||(T_{\rm ideal})_{11}-(T_{\rm expt.})_{11} ||^2+||(T_{\rm ideal})_{12}-(T_{\rm expt.})_{12} ||^2\\
&+||(T_{\rm ideal})_{21}-(T_{\rm expt.})_{21}  ||^2+|| (T_{\rm ideal})_{22}-(T_{\rm expt.})_{22}||^2
\end{split}
\end{equation}
with $T_{11}=1$, $T_{12}=\vec{0}$, $T_{21}=\vec{t}$, and $T_{22}=R$ in Eq.~\ref{eq:PTM_matrix}~.
Markovian noise affects populations and expectation values independently of coherent rotations and therefore manifests itself in the first row and first column of the PTM. We therefore define the Markovian error contribution as the squared Frobenius distance between the corresponding first row and first column entries of the ideal and experimental PTMs:

\begin{equation}
\begin{split}
   & \epsilon_{\rm Mark.}^2 = ||(T_{\rm ideal})_{11}-(T_{\rm expt.})_{11}||^2 \\
   &+ \sum_{j=2,3,4} ||(T_{\rm ideal})_{1j}-(T_{\rm expt.})_{1j}||^2+ \sum_{j=2,3,4} ||(T_{\rm ideal})_{j1}-(T_{\rm expt.})_{j1}||^2 
\end{split}
\end{equation}
\subsubsection{Coherent errors}
Once the Markovian contribution is fixed by the first row and column of the PTM in Eq.(\ref{eq:PTM_matrix}), the remaining deviation between the ideal and experimental maps is encoded in the 
$3\times 3$ Bloch-sphere vector, quantified by the Frobenius distance between $3\times 3$ blocks. In particular, the total error can be written as 
\begin{equation}\label{eq:total_error}
\epsilon_{\rm total}^2 = \epsilon_{\rm Mark.}^2 + ||R_{\rm ideal}-R_{\rm expt.}||^2
\end{equation}
where $R_{\rm ideal}$ and $R_{\rm expt.}$ are the  $3\times 3$ sub-blocks of $T_{\rm ideal}$ and $T_{\rm expt.}$, respectively. The matrix $R$ describes the linear action of the map on the Bloch vector. For an ideal unitary map, this action is a rigid rotation of the Bloch vector, hence $R_{\rm ideal} \in SO(3)$. Using the orthogonality of $R_{\rm ideal}$, we may write

\begin{equation}\label{eq:orthogonality_R}
||R_{\rm ideal}-R_{\rm expt.}||^2 = ||I-R_{\rm ideal}^T R_{\rm expt.}||^2.
\end{equation}
Introducing the relative transformation,

\begin{equation}\label{eq:M}
M=R_{\rm ideal}^T R_{\rm expt.},
\end{equation}
removes the ideal rotation, leaving a transformation that captures only the experimental error.

If the experimental imperfections are purely coherent, then the implemented operation differs from the ideal one only by an additional unitary rotation. In the PTM picture that implies that $R_{\rm expt.}$ remains approximately orthogonal, and therefore $M$ itself is an orthogonal matrix describing a residual rotation. Coherent errors therefore correspond to the orthogonal part of $M$, i.e., to deviations that preserve the Bloch vector length and its angles. 

In the next subsection we will describe non-Markovian contributions that can not be described by either of the mechanisms outlined above.

\subsubsection{Non-Markovian errors}

Having identified (i) Markovian errors via the first row and column of the PTM and (ii) coherent errors as residual Bloch-sphere rotations, we now define the non-Markovian contributions as the remaining deviations that cannot be explained by either of these mechanisms. Physically, such errors arise from temporal correlations, slow drifts, or environmental memory effects, and appear as non-orthogonal deformations of the Bloch sphere, i.e., stretching or squeezing rather than a rigid rotation of this.\\

To separate rotations from deformations, we factor $M$ into orthogonal and symmetric components. We shall show below that this is always possible. Using a right-polar decomposition, we write

\begin{equation}\label{eq:QR}
M=PR,
\end{equation}
where $P$ is symmetric, positive-semi-definite and $R$ is orthogonal. If the error is completely coherent, then $M$ is orthogonal and $P=I$. However, if the error is fully non-Markovian, then $R=I$ and $M$ is symmetric.
The physical interpretation of this decomposition follows from the fact that any coherent miscalibration corresponds to a unitary error and therefore induces a rigid, length-preserving rotation of the Bloch sphere, represented by an orthogonal matrix. Consequently, if the experimental deviation from the ideal operation is purely coherent, the transformation $M$ is orthogonal and the symmetric component is trivial, $P=1$. Conversely, deviations that cannot be captured by a single unitary rotation necessarily correspond to non-unitary dynamics, such as those arising from temporal fluctuations or memory effects. After factoring out the closest orthogonal transformation, such effects manifest as symmetric deformations of the Bloch sphere encoded entirely in $P$, so that in the absence of coherent errors one has $R=1$. 
We then define the coherent error magnitude as

\begin{equation}\label{eq:coh_error}
\epsilon_{\rm Cohr.}^2 =  ||1-R||^2,
\end{equation}

which quantifies the size of the residual rotation due to unitary miscalibrations, and the non-Markovian error as

\begin{equation}\label{eq:nonMark_error}
\epsilon_{\rm NonMark.}^2 =  ||1-P||^2
\end{equation}
which quantifies the size of the symmetric deformation, i.e., the non-unitary distortions that are associated with non-Markovian effects.

\subsection{Small error additivity}
In general the orthogonal and symmetric elements do not contribute additively to $||I-M||^2$. In particular, we have

\begin{equation}\label{eq:nonaddit}
||I-M||^2 \neq  \epsilon_{\rm NonMark.}^2+ \epsilon_{\rm Cohr.}^2
\end{equation}
This reflects the fact that, for finite errors, the orthogonal and symmetric parts of $M$ are not independent and may contribute cross terms to the Frobenious norm.
However, in the small-error regime, a perturbative expansion shows that these contributions become approximately additive (See Eq.~17-20 for clarification). 
To analyze the additivity of errors in the small-regime,  we note that this norm is invariant under sign changes, so that $||I-M||=||M-I||$. We therefore expand the deviation $M-I$, which represents small errors as perturbations about the identity transformation. Any matrix can be uniquely decomposed into symmetric  and antisymmetric components~\cite{hoffman1971linear, strang2023introduction}, and applying this decomposition to $I-M$ yields:

\begin{equation}
\begin{split}
I-M &=\frac{1}{2}[(I-M)+(I-M)^T]+\frac{1}{2}[(M-I)-(M-I)^T]\\
&=\frac{1}{2}(M+M^T -2I) +  \frac{1}{2}(M-M^T),
\end{split}
\end{equation}
i.e., the sum of a symmetric and an antisymmetric component. Since these two components are othogonal under the Frobenious  inner product, it follows that

\begin{equation}
||I-M||^2 =  ||\frac{1}{2}(M+M^T -2I)||^2 +  ||\frac{1}{2}(M-M^T)||^2
\end{equation}

In the small error limit, these components admit a direct physical interpretation. Specifically, the antisymmetric part of $M$ generates the orthogonal matrix $R$~\cite{hoffman1971linear, strang2023introduction} to leading order,

\begin{equation}
R\approx I + \frac{1}{2}(M-M^T)
\end{equation}
while the symmetric part generates the matrix $Q$,

\begin{equation}
Q \approx  \frac{1}{2}(M+M^T).
\end{equation}
 Hence,

\begin{equation}
||I-M||^2 =  ||I-P||^2 +  ||I-R||^2 \approx  \epsilon_{\rm NonMark.}^2+ \epsilon_{\rm Cohr.}^2 
\end{equation}
Adding back the Markovian contribution from Eq.(\ref{eq:total_error}) then yields the full error decomposition

\begin{equation}
\epsilon_{\rm total}^2 \approx \epsilon_{\rm Mark.}^2 +\epsilon_{\rm NonMark.}^2 + \epsilon_{\rm Cohr.}^2.
\end{equation}

\subsection{Relation to entanglement fidelity and gate fidelity}

We now relate this description to a commonly used figure of merit in quantum information experiments, namely the entanglement fidelity \cite{PhysRevA.54.2614}.  The entanglement fidelity provides a natural measure of how closely an implemented quantum process $\Phi$ approximates a target unitary gate $U$. In the Choi matrix representation, the entanglement fidelity between a unitary gate $U$ and a quantum map $\Phi$ is given by \cite{PhysRevA.60.1888,NIELSEN2002249} 

\begin{equation}
F_e=\frac{1}{4}{\rm Tr}[\rho^{U}\rho^{\Phi}].
\end{equation}
For a single qubit ($d=2$), the entanglement fidelity is directly related to the averaged gate fidelity~\cite{NIELSEN2002249} by

\begin{equation}\label{eq:F_gate}
    F_{av} = \frac{d F_{e}+1}{d+1} = \frac{2F_{e}+1}{3}.
\end{equation}

While the Choi matrix provides the most direct definition of the entanglement fidelity, it is often more convenient to express this quantity in the PTM representation. Owing to the linear equivalence between the Choi and PTM representations (see Appendix \ref{app:CMtoPTM}), 
the entanglement fidelity can be written as

\begin{equation}\label{eq:Fe_PTM_1}
    F_e = \frac{1}{4}\text{Tr}[T^{U} (T^{\Phi})^T]
\end{equation}
where $T^{U}=T_{\rm ideal}$ and $T^{\Phi}=T_{\rm expt.}$ denote the PTMs of the unitary target operation and of the implemented map, respectively.

We now contrast this fidelity-based metric with the distance-based error measured introduced above. In the PTM representation we define the total error between the ideal and experimental implementations as the squared Frobenius distance between their respective PTMs, i.e.,

\begin{equation}\label{eq:total_err_equiv}
\begin{split}
    \epsilon_{\rm total}^2 &\equiv||T_{\rm ideal}-T_{\rm expt.}||^2 \\
    &= ||T_{\rm ideal}||^2 + ||T_{\rm expt.}||^2 - 2 \text{Tr}[T_{\rm ideal} (T_{\rm expt.})^T]
    \end{split}
\end{equation}
The last term in the expression above is directly proportional to the entanglement fidelity $F_e$ (Eq.~\ref{eq:Fe_PTM_1}).
Substituting Eq.~(\ref{eq:Fe_PTM_1}) into Eq.~\ref{eq:total_err_equiv} we obtain

\begin{equation}\label{eq:total_err}
\begin{split}
    \epsilon_{\rm total}^2 \approx 8(1-F_e),
    \end{split}
\end{equation}
where we have used $||T_{\rm ideal}||^2=4$ for a single-qubit unitary, and assumed that the experimental channel is close to the unitary  $||T_{\rm expt.}||^2=4 +O(\epsilon^2)$, with higher-order terms being negligible. Using the relation of the entanglement fidelity with the gate fidelity from Eq.~\ref{eq:F_gate} for $d=2$, we can then relate the total error to the gate infidelity, according to

\begin{equation}\label{eq:total_err_2}
\begin{split}
    r\equiv  1-F_{av}\approx\frac{1}{12}\epsilon_{\rm total}^2 
    \end{split}
\end{equation}
We emphasize that this relation holds only in the perturbative regime.
Now the total error $\epsilon_{\rm total}^2$ is of course nonzero whenever the gate fidelity is less than unity. However, these two quantities exhibit different sensitivities to specific error mechanisms.  In particular, the Frobenius error is first-order sensitive to coherent errors and to certain Markovian contributions, whereas the gate fidelity is generally second-order sensitive to small coherent miscalibrations. This distinction highlights an important tradeoff between distance-based and fidelity-based error metrics. While the average gate fidelity suppresses first-order coherent errors, making it robust to small miscalibrations, the Frobenius-distance metric employed in this work remains linearly sensitive to such errors. As a result, the tangent-space decomposition enables the detection of coherent and non-Markovian error mechanisms that may remain invisible in fidelity-based benchmarks. 




%
\section{Experimental system and results}\label{Sec:experiment}

\subsection{Tomography protocol for a single-qubit}\label{subsec:rcm}
The tomography protocol for a single qubit is summarized by the circuit in Fig.~\ref{Fig1}(c). The qubit is prepared in the state $\ket{0}$, a known gate $A$ is applied first, followed by the trace-preserving Pauli channel $T_{\rm expt.}$, and then another known gate $B$ is applied before measuring the population in the state $\ket{1}$.
In the PTM representation, the effective process implemented by this circuit is the composition $T_B T_{expt} T_A$, where $T_A$ and $T_B$ are the PTMs of the known gates $A$ and $B$, and $T_{\rm expt.}$ is the PTM that we wish to characterize.

Let $\vec{p}_{\ket{0}}=(1/2, 0, 0, 1/2)^T$ and $\vec{p}_{\ket{1}}(1/2, 0, 0, -1/2)^T$ denote the Pauli vectors associated with the input state $\ket{0}\bra{0}$ and the measurement effect $\ket{1}\bra{1}$, respectively. The expected outcome of this measurement is then given by

\begin{equation}\label{eq:y_1}
\begin{split}
    y =& \vec{p}^T_{\ket{1}} T_B T_{\rm expt.} T_A \vec{p}_{\ket{0}} = \text{Tr}[\vec{p}_{\ket{0} }
 \vec{p}^T_{\ket{1}} T_B T_{\rm expt.} T_A ] \\
 =& \text{Tr}[ T_A \vec{p}_{\ket{0} }
 \vec{p}^T_{\ket{1}} T_B T_{\rm expt.}  ] = \text{Tr}[Y T_{\rm expt.}, ]
 \end{split}
\end{equation}
with $Y = T_A \vec{p}_{\ket{0} } \vec{p}^T_{\ket{1}} T_B $.
Here, to express the scalar as a trace, we have used the identity $uXv={\rm Tr}[Xvu^T]$ (valid for any vectors $u$, $v$ and matrix $X$), and the cyclic property of the trace.

Repeating this procedure for a set of linearly independent set of choices of gates ($A$,$B$) produces measurement outcomes $\{ y_i\}$ that are linear functions of the unknown PTM $T_{\rm expt.}$ that we want to reconstruct and analyze. Denoting the corresponding gates by $(A_i, B_i)$, and defining the corresponding $Y_i = T_{A_i} \vec{p}_{\ket{0} }
 \vec{p}^T_{\ket{1}} T_{B_i}$, we then obtain the linear measurement model

 \begin{equation}\label{eq:y_2}
     y_i = \text{Tr}[Y_i T_{\rm expt.}],
 \end{equation}
where $y_i$ denotes measurement outcome $i$, and $Y_i$ the effective composite channel for this outcome.
In principle, one can reconstruct $T_{\rm expt.}$ by solving the  set of linear equations expressed by Eq.~\ref{eq:y_2}. However, in practice, to improve robustness to experimental noise, it is more effective to estimate $T_{\rm expt.}$ by minimizing a least-squares cost function. Explicitly, with
$y_i$ the experimental measurement outcomes, we minimize the function
 
 \begin{equation}\label{eq:y_3}
     f(\tilde{T}) = \sum_i | \text{Tr}[Y_i \tilde{T}]-y_i|^2,
 \end{equation}

 over all real $4\times 4$ matrices $\tilde{T}$. This results in optimization of 16 parameters.
 
 There are several additional conditions that provide constraints which reduce the number of parameters and lead to an efficiently solvable formulation of the optimization as a convex optimization. First, trace preservation imposes linear constraints on the PTM. For a trace-preserving map, the first row of the PTM is fixed to $(1,0,0,0)$, or equivalently, $T_{0\beta}=0$ for $\beta\neq 0$ in Eq.(\ref{eq:Choi_PTM}). This reduces the number of real parameters from 16 to 12, consistent with the dimension of the trace-preserving manifold discussed in Sec.~\ref{Sec:tangent}. 
 
 Second, recall that the PTM representation makes the measurement model linear, as shown in Eq.~\ref{eq:y_1}-\ref{eq:y_3}, while complete positivity is most naturally enforced in the Choi representation, where it corresponds to a semidefinite constraint. We can therefore impose physicality by constraining the Choi matrix associated with $\tilde{T}$ to be positive semidefinite.

Finally, complete positivity of a map $\Phi$ implies positivity of its Choi matrix. Writing the Choi matrix in the Pauli basis (see Eq.~\ref{eq:Choi_PTM})) then leads to the semidefinite constraint 

\begin{equation}\label{PSD}
     \sum_{\alpha,\beta} \tilde{T}_{\alpha\beta}\sigma_{\alpha}\otimes \sigma_{\beta}^T \succeq 0.
 \end{equation}
 Unlike the trace-preserving constraints, Eq.~\ref{PSD} is not linear in the entries of $\tilde{T}_{\alpha, \beta}$. 
 Instead, it expresses a convex optimization problem that can be solved using standard semidefinite programming (SDP) techniques.


 


\subsection{A single trapped ion experiment}\label{subsec:trappedion_expt} 

\begin{figure}
    \centering
    \includegraphics[width=0.8\linewidth]{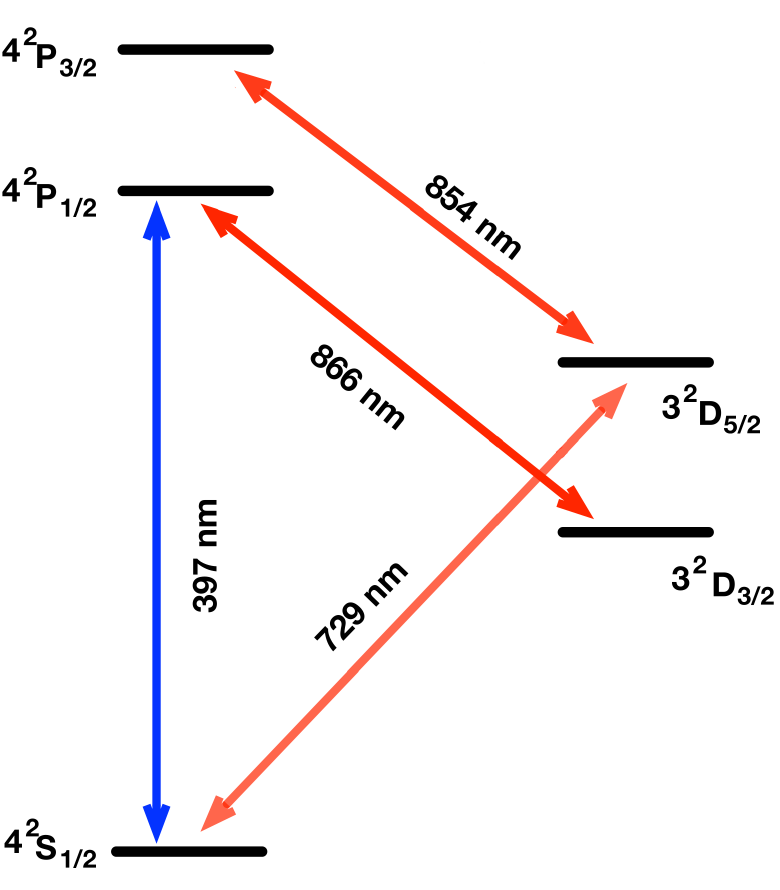}
    \caption{Energy level scheme of the Ca$^+$ ion showing the main transitions used for the experiment.}
    \label{fig:ca+ion}
\end{figure}

Here we apply the RCM protocol to a qubit implemented within the electronic states of a trapped ion.
In this case, the intended unitary operation $U$ (see Fig.~\ref{Fig1}) under characterization is a $\pi$ rotation around the $x$-axis, i.e. $U\equiv R_x(\pi)=\exp(-i\pi\sigma_{x}/2)$. 

A single $^{40}\text{Ca}^+$ ion is stored in a surface-electrode radiofrequency ion trap characterized by secular axial and radial motion frequencies: $(\omega_{ax}, \omega_{r1}, \omega_{r2}) = (1.6, 3.4, 3.6)\,\text{MHz}$.
%
The transition between the ground state $\text{S}_{1/2}$ and the metastable state $\text{D}_{5/2}$ (see Fig.~\ref{fig:ca+ion}) is driven with laser light at $729\,\text{nm}$ generated by an extended cavity diode laser, stabilized by locking onto an ultra-low expansion (ULE) reference cavity, and amplified through multiple stages to increase its power. Using approximately $4\,\text{mW}$ of optical power, we achieve Rabi oscillations between the $\text{S}$ and $\text{D}$ states with a frequency close to $250\,\text{kHz}$.
%
A static magnetic field of approximately $4.3\,\text{G}$ is applied using permanent magnets along the axial trapping direction. This lifts the degeneracy of the $10$ transitions between the Zeeman states of the $\text{S}_{1/2}-\text{D}_{5/2}$ manifold. The geometric arrangement of the magnetic field, the direction of the laser beam, and its polarization prevent the excitation of transitions with a change in the magnetic quantum number of $\Delta m_j = \pm 1$.
%
The qubit is an optical qubit encoded in the Zeeman states $\ket{S} \equiv \text{S}_{1/2,\,m = -1/2}$ of the ground state and $\ket{D} \equiv \text{D}_{5/2,\,m' = -1/2}$ of the metastable state (lifetime $\sim 1\,\text{s}$). Single-qubit operations are performed by driving the quadrupole transition with laser pulses at $729\,\text{nm}$.

In order to enable determination of the $12$ independent parameters of the experimental PTM $T_{expt.}$, we carry out $12$ different experimental sequences. Each of the $12$ experimental sequences used in the RCM protocol consists of the following steps:
\begin{itemize}
\item[(a)] Doppler Cooling: The ion is Doppler cooled for $200\,\mu\text{s}$ on the $\text{S}_{1/2} \leftrightarrow \text{P}_{1/2}$ transition at $397\,\text{nm}$. At the same time, repump lasers at $866\,\text{nm}$ and $854\,\text{nm}$ are used to prevent optical pumping into the dark $D$ states. See Fig.~\ref{fig:ca+ion} for the energy level scheme of the Ca$^+$ ion.
\item[(b)] Optical Pumping: The ion is optically pumped for approximately $1\,\text{ms}$ into the $\ket{S}$ state by driving the transition $\text{S}_{1/2,\,m = 1/2} \leftrightarrow \text{D}_{5/2,\,m = -3/2}$.  
\item[(c)] Sideband Cooling: Motion along the axial mode is reduced with $6$ cycles of sideband cooling, each lasting $200\,\mu\text{s}$. Sideband cooling is performed on the $\text{S}_{1/2,\,m = -1/2} \leftrightarrow \text{D}_{5/2,\,m = -5/2}$ transition. After each cycle, optical pumping is performed as described in step (b).
\item[(d)] Quantum Operations: The core part of the experimental sequence comprises up to three operations, as illustrated in Fig.~\ref{Fig1}. In these experiments, our target unitary gate is a $R_x(\pi)$ rotation, $U =R_x(\pi)$ and we wish to measure the corresponding PTM $T_{expt.}$. The gate sequence in Fig.~\ref{Fig1} (c) is given by $B U A$, with $A$ and $B$ specified as qubit pre- and post-rotations $R_1$ and $R_2$ that are selected for each of the $12$ experimental sequences chosen from the two sets $A\in \{ I, R_x(\pi), R_x(\pi/2), R_y(\pi)\}$ and $B\in \{I, R_y(\pi/2), R_x(\pi/2)\}$.

Experimentally, these rotations are realized by applying resonant laser pulses on the qubit transition for a duration $t=\pi/(N\Omega)$, where $\Omega$ denotes the Rabi frequency and $N=1$ for $R_x(\pi)$ ($t(N=1)=\pi$-pulse time), while $N=2$ for $R_x(\pi/2)$ and $R_y(\pi/2)$. The rotation axis is defined by the laser phase: $\phi=0$ corresponds to rotations about the x-axis, and $\phi=\pi/2$ to rotations about the y-axis. As a consequence, $R_1$ and $R_2$ may have different durations, i.e., a full $\pi$-pulse time, a half $\pi$-pulse time, or zero in the case of the identity. To standardize the duration of each operation to a full $\pi$-pulse time, a padding time with no laser excitation is added after the shorter operations. For the identity operation, the qubit is simply left to evolve freely for a $\pi$-pulse time.

\item[(e)] State Readout: The internal state of the ion is measured by using the electron shelving technique. To determine whether the ion has transitioned to the $\ket{D}$ state, we apply laser light at $397\,\text{nm}$ and $866\,\text{nm}$ and collect the ion's fluorescence at $397\,\text{nm}$ using a photomultiplier tube. The ion scatters blue light only if it is in the $\ket{S}$ state, enabling reliable discrimination between the two qubit states. 
\end{itemize}

This sequence is repeated $10^4$ times from (a) to (e) to measure the probability $\mathcal{P}_D$ of the ion to be in the $\ket{D}$ qubit state.
%
Before each set of $10^4$ runs for each of the $12$ tomography measurements, a sequence of scripted calibrations is performed to establish the best estimate for some key experimental parameters of the sequence for that specific tomography measurement, such as the $\pi$-pulse time.
First, carrier spectroscopy is carried out on the $\text{S}_{1/2}-\text{D}_{1/2}$ and $\text{S}_{1/2}-\text{D}_{5/2}$ transitions to determine the frequency shift required to maintain the $729\,\text{nm}$ laser resonant with the qubit transition as well as the other transitions used in that experimental sequence. Second, the measured frequency shift of the qubit transition is recorded and tracked over time, enabling the prediction of frequency drifts via linear extrapolation and compensation during the experimental runs.
Third, the axial secular frequency is determined by driving the first axial blue sideband of the qubit transition $\text{S}_{1/2,\,m = -1/2} \leftrightarrow \text{D}_{5/2,\,m = -5/2}$. Knowledge of the axial mode frequency is essential for calibrating the parameters used in sideband cooling (step (c) of the experimental sequence above). In particular, the power of the $854\,\text{nm}$ repump laser is adjusted to minimize the residual population in the $D$ state, while the number of cooling cycles and the pulse duration are kept fixed.
Finally, the $\pi$-pulse time is calibrated by performing a Rabi oscillation measurement on the carrier transition. The measured population dynamics are fitted to a squared sinusoidal model to accurately extract the $\pi$-pulse time.

\subsection{Inducing Coherent Errors}\label{sec:coherent}

\begin{figure*}
    \centering
    \includegraphics[width=\linewidth]{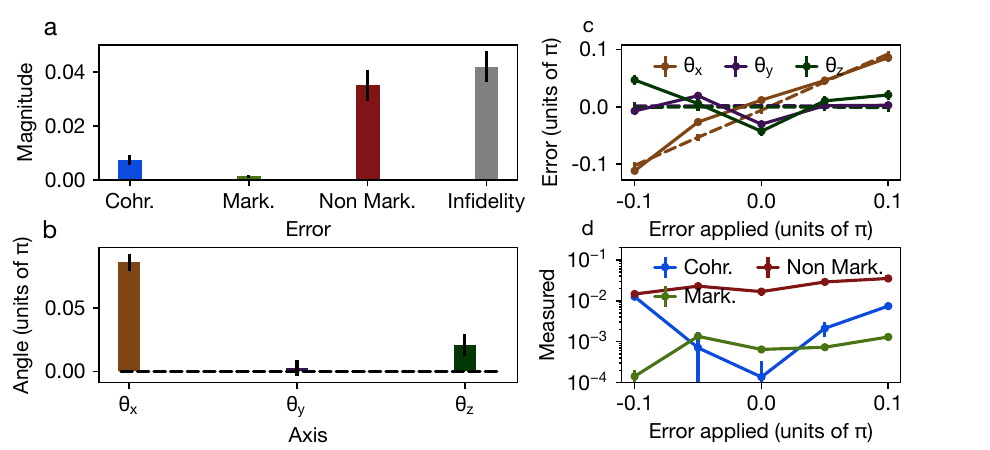}
    \caption{Process tomography for a single qubit rotation with induced coherent errors. (a)  Contributions of different types of error $\epsilon^2$ to the total gate infidelity $r$ when a $\pi$ rotation about the $x$ axis, $U= R_x(\pi/2)$, is miscalibrated by 10\% under-rotation, i.e., the miscalibrated $\pi$-pulse time is $0.9$ of the true $\pi$-pulse time. A non-Markovian contribution arising from background fluctuations is also found to be present. (b) Rotation-angle miscalibration inferred from the tangent-space decomposition: the angle estimates are extracted via he axis-angle representation (Eq.~\ref{ax-angle}). (c) Measured rotation-angle error (y-axis) as a function of the introduced $\pi$-pulse time miscalibration for the rotation about the $x$-axis. Solid lines show experimental data, dashed lines show numerical simulations (see Appendix ~\ref{app:numerical_sim}). Error bars for the histograms in (a) and (b) are indicated by the vertical black lines. (d) Measured total errors. While the non-Markovian errors are constant, the coherent errors show the expected parabolic structure (see main text).}
    \label{fig:coherent_errors}
\end{figure*}

To test the tangent-space decomposition theory and identify the physical origin of the errors for the single qubit gate, we deliberately introduce a coherent error realized as over- or under-rotations of the unitary operation $U$ under study, while no deliberate errors are introduced in any of the other tomography operations. 
Throughout the RCM protocol, we restrict ourselves to the carrier qubit transition $\text{S}_{1/2,\,m = -1/2} \leftrightarrow \text{D}_{5/2,\,m = -5/2}$, such that the laser couples only the internal spin states and the phonon number remains conserved. 

An on-resonance rotation about the $x$-axis is described by the matrix

\begin{equation}
\begin{split}
R_x(\theta)=e^{-i \theta X / 2}=
&\begin{pmatrix}
\cos (\frac{\theta}{2} ) & -i\sin(\frac{\theta}{2})\\
-i\sin(\frac{\theta}{2}) & \cos(\frac{\theta}{2})
\end{pmatrix}
\end{split},
\end{equation}
with $\theta=\Omega t$ the pulse area (rotation angle), where $\Omega$ is the Rabi frequency and $t$ is the $\pi$-pulse time described in Sec.~\ref{subsec:trappedion_expt}.
Introducing a systematic coherent error means modifying the rotation angle to

\begin{equation}
    \theta\rightarrow\theta(1+\varepsilon)
\end{equation}

where $\varepsilon$ denotes the fractional over- or under- rotation, i.e., a deterministic miscalibration that remains constant from shot to shot and throughout the entire set of $12$ tomography measurements. 
The rotation angle miscalibration can be readily implemented by choosing an incorrect pulse duration, i.e., the $\pi$-pulse time $t$ is modified to $t(1+\varepsilon)$, with $\Omega$ fixed. 
We use the $12$ measurement settings in the tomography sequence, described above, including all possible combinations of $A$ and $B$. This produces $12$ data points $y_1, \cdots, y_{12}$ for $12$ values $Y_1, \cdots, Y_{12}$ in Eq.~\ref{eq:y_2}. We then use CVXPY --- a PYTHON package for semi-definite programming --- to minimize the cost function $f$ in Eq.~\ref{eq:y_3} under the inequality constraint Eq.~\ref{PSD}. This results in the PTM that represents a completely positive map and minimizes $f$, which is then mapped to a Choi Matrix
(see Eq.~\ref{eq:y_3}). 
A similar technique is used to correct for measurement setting errors (see Appendix B)  The measured Choi matrix (or PTM) is used to extract the experimental rotation operator $R_{\text{expt.}}$. Following the right-polar  decomposition (Eq.~\ref{eq:QR}), the coherent error is then represented by Eq.~\ref{eq:coh_error}. 
One can further use the rotation matrix to go beyond estimation of only the magnitude of the coherent error, and also estimate the rotation angle with respect to each axis. That is, we may write $R$ in the axis-angle representation as $R = R_{\hat{n}}(\delta \theta)$. We then define 
\begin{equation}\label{ax-angle}
    \delta \theta_x = \hat{n}_x\delta\theta,\ \  \delta \theta_y = \hat{n}_y\delta\theta\  \text{ and }\  \delta \theta_z = \hat{n}_z\delta\theta.
\end{equation}
This reveals (up to commutative ambiguities) the angle of rotations about each axis in the error.

Fig.~\ref{fig:coherent_errors}(a) shows the results of the tangent-space decomposition for a single-qubit $\pi$ rotation about the $x$ axis that is subjected to a $10\%$ under-rotation. 
The inferred rotation about the $x$ axis correctly matches the introduced error as shown in Fig.~\ref{fig:coherent_errors}(b), whereas the considerably smaller values of rotation inferred about the other two axes are not statistically significant, since their values remain within the corresponding error bars. 

Fig.~\ref{fig:coherent_errors}(c) shows the rotation angles about the three axes inferred from the tangent-space decompositions as functions of the applied miscalibration angle of rotation about the $x$ axis, ranging from a $0.9 \pi$ under-rotation to $1.1 \pi$ over-rotation, together with theoretical simulation results for comparison.
The introduced miscalibrations are chosen to be small perturbations (up to $10\%$) relative to the total rotation, since the tangent-space decomposition is valid only in the linear regime, i.e. for small perturbations. 

In Fig.~\ref{fig:coherent_errors}(d) the measured errors are plotted as a function of the introduced miscalibration. The non-Markovian noise contribution to the total error remains approximately constant over the full range of miscalibration values, indicating that it is a background noise source independent of the applied miscalibration. In contrast, the coherent noise exhibits a parabolic dependency, as expected: in fact for a small deviation, $\delta\theta$ from the ideal case, the fidelity between the ideal and the real state is given by $\cos^2(\delta\theta) \approx 1 - (\delta\theta)^2$. Therefore the error, i.e., infidelity $\epsilon_{\text{Cohr.}}^2$ is quadratic in $\delta\theta$.

Overall, we observe very good agreement between experimental data and simulation for rotations about the $x$ axis when the coherent error is deliberately introduced. 
This demonstrates the ability of the tangent-space decomposition method to correctly detect and identify coherent errors, even in the presence of native and not-negligible background non-Markovian noise. This is particularly evident in Fig.~\ref{fig:coherent_errors}(c), where induced miscalibrations at $\pm\,0.05 \pi$ result in a coherent error that is an order of magnitude smaller than the non-Markovian noise contribution.

\subsection{Inducing Non-Markovian noise}\label{sec:nonMarkov}

\begin{figure}
    \centering
    \includegraphics[width=\linewidth]{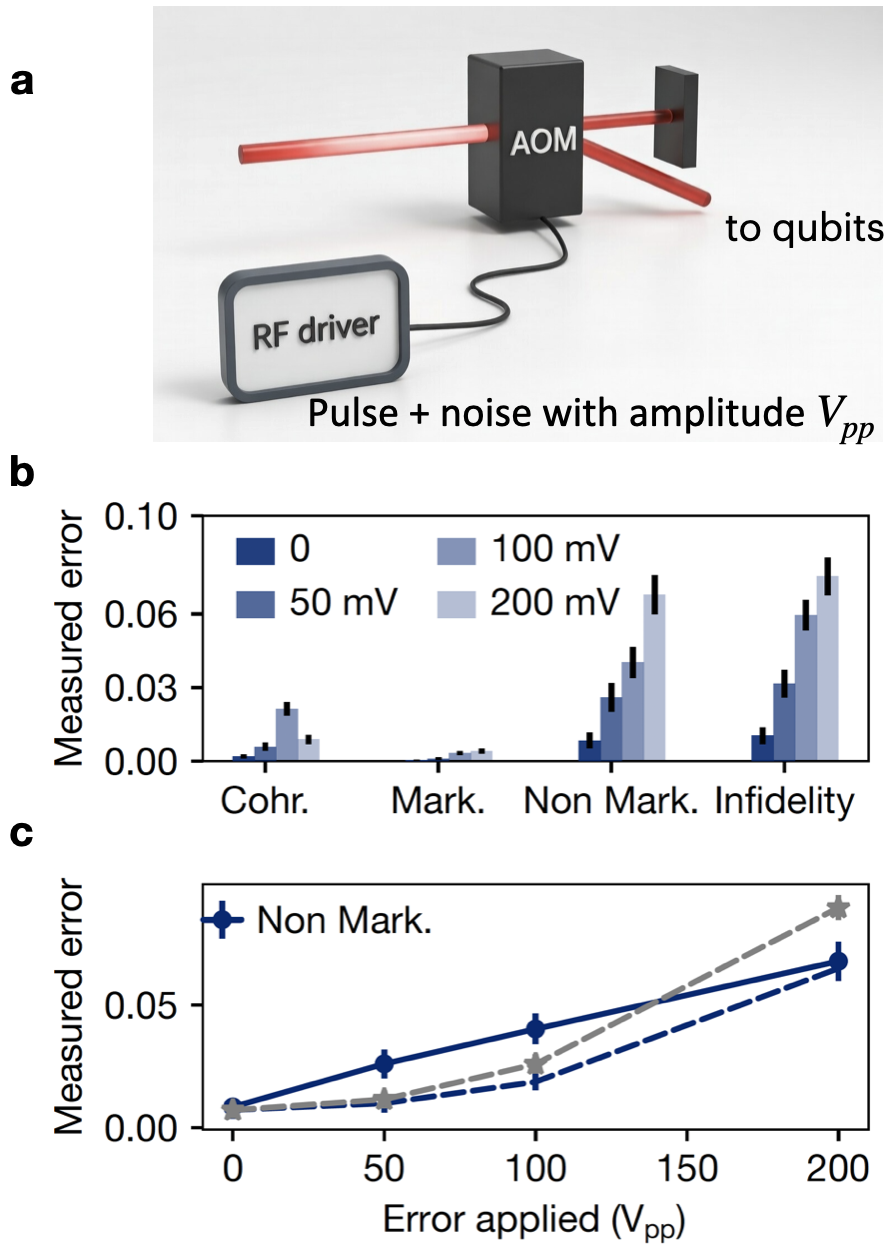}
    \caption{
    (a) Non-Markovian noise is emulated by introducing controlled fluctuations in the Rabi frequency via laser power modulation, which is implemented by modulating the RF signal fed on a Acousto-Optic Modulator (AOM). 
    (b) Contributions of the different errors to the total gate infidelity as a function of different RF drive amplitudes $V_{pp}$. Measured non-Markovian errors and contribution to the infidelity monotonically increases with $V_{pp}$. 
    (c) Measured total error ($y$-axis) versus injected $V_{pp}$ RF amplitude ($x$-axis). Solid lines show experimental data; dashed lines show numerical simulations (see Appendix ~\ref{app:numerical_sim}). Blue (gray) dashed lines corresponds to normal (top hat) error distribution.
}
    \label{fig:non-Markovian}
\end{figure}

As described in detail in Sec.~\ref{subsec:trappedion_expt}, single-qubit rotations are implemented by driving the quadrupole transition with laser light at 729~nm. 
The laser frequency is precisely tuned to the qubit resonance using an Acousto-Optic Modulator (AOM), which is driven by a radiofrequency (RF) signal, as shown in Fig.~\ref{fig:non-Markovian}(a). 
The optical power delivered to the qubit depends on RF drive power applied to the AOM. Consequently, the laser power and hence also the Rabi frequency can be controlled by adjusting the RF signal amplitude. 
To introduce artificial non-Markovian noise into the system, we apply an amplitude modulation to the RF drive at a modulation frequency of $10$~kHz, with varying amplitudes that are sufficiently small to be treated as perturbations, resulting in perturbations of the modulation depth.  
The applied modulation frequency gives rise to slow fluctuations of the rotation angle $\theta$ that can be considered constant over the duration of an individual experimental shot, while varying from shot-to-shot, yielding a normal distribution $\mathcal{N}$ for large numbers of shots

\begin{equation}
    \label{eq:non-mark}
    \theta \rightarrow \theta(1+\delta ), \hspace{0.1cm} \delta \sim \mathcal{N}(0, \sigma^{2}).
\end{equation}

Experimental results with such induced non-Markovian noise are reported in Fig.~\ref{fig:non-Markovian}(b). The non-Markovian contribution to the total infidelity extracted via the tangent-space decomposition method is seen to exhibit a monotonic dependence on the amplitude of the RF drive used for the modulation. 
We expect a linear dependence, since the perturbation is sufficiently small to remain within the linear-response regime, where the fidelity varies linearly with the fluctuations of the rotation angle $\theta$ as described by Eq.~\ref{eq:non-mark}.

The non-Markovian noise contribution detected by the tangent-space decomposition method is fundamentally sensitive to variance of the underlying noise distribution governing the fluctuations of the control parameters (e.g., $\sigma^{2}$ for the normal distribution of Eq.~\ref{eq:non-mark}. 
Figures~\ref{fig:non-Markovian}~(b,~c) show that the extracted non-Markovian component exhibits a monotonic and approximately linear dependence on the induced noise strength in the small-perturbation regime.
However, the way in which the native (background) non-Markovian noise - also detected in the experiments reported in Figures~\ref{fig:coherent_errors}~(b,~c) - combines with the deliberately injected non-Markovian noise is not directly accessible and experimentally unknown.
The method therefore detects and isolates the presence of the overall non-Markovian contributions at the experimental level, without requiring detailed knowledge of their physical origin or statistical distribution, making it a robust and practically useful operative tool.

We also investigated the role of different noise distributions in the  (see Appendix ~\ref{app:numerical_sim}), comparing normal and "top hat" models for the injected fluctuations, as shown in Fig.~\ref{fig:coherent_errors}~(c). 
Within the sensitivity and resolution of the experimental data, no statistically significant distinction between these distributions could be resolved.
While the tangent-space decomposition framework is distribution-agnostic and robustly isolates non-Markovian noise contributions independently of the specific form of the noise distribution, it does not by itself allow for a full reconstruction of the noise statistics without additional constraints or measurements.

\section{Summary and Outlook}
In this work, we have presented a geometry-based framework for diagnosing coherent, Markovian, and non-Markovian errors in unitary quantum operations using a tangent-space decomposition on the manifold of CPTP maps. The framework combines the Choi matrix representation for CPTP maps with the Pauli Transfer Matrix, PTM, representation. This allows the measurement model and error decomposition to be expressed in the PTM representation, while taking advantage of complete positivity constraints on the Choi matrix to construct a robust and effective tomographic technique to characterize multiple sources of error. 
We applied this framework to experimental study of single-qubit gates implemented in a trapped $^{40}\mathrm{Ca}^+$ ion system.
By deliberately introducing controlled coherent
miscalibrations and engineered non-Markovian fluctuations,
we demonstrated that the tangent-space decomposition
correctly identifies the physical origin of each contribution.
In particular, deterministic over- and under-rotations were
faithfully extracted as orthogonal (unitary) components of the
error, while slow shot-to-shot fluctuations manifested as
symmetric deformations of the Bloch sphere. The extracted
error components show strong agreement with numerical
simulations, confirming that the geometric decomposition
accurately captures the underlying structure of experimentally
relevant noise processes.

The key advantage of this approach is that it
provides a single-snapshot diagnostic tool, without requiring multi-time tomography. 
Thus, it does not assume a specific Lindblad generator, noise
spectrum, or temporal kernel, yet it isolates physically
distinct error mechanisms directly from tomographic data.
While the method does not reconstruct the full multi-time
memory structure of the environment, it provides an
operationally accessible quantification of coherent and
non-unitary contributions at the gate level, complementing
multi-time process-tensor or noise-spectroscopy approaches.
Integrating the tangent-space decomposition
with adaptive calibration routines or feedback-based control
protocols can enable selective suppression of distinct error
mechanisms. 

The tangent space decomposition can be scaled up to make it practical for experiments involving larger number of qubits, using an explicit tomographic procedure that we term a Reduced Choi Matrix (RCM) protocol~\cite{madhusudhana2023benchmarkingmultiqubitgates, madhusudhana2023benchmarkingmultiqubitgates2}, that retains full sensitivity to coherent and non-Markovian contributions, while requiring significantly fewer measurements than full process tomography. 
In RCM tomography, one can efficiently obtain the reduced Choi-matrix corresponding to one and two-qubit subsystems of an $N$ qubit system. The Choi-matrices corresponding to all $k-$qubit subsystems can be obtained with $\mathcal O(N^k 16^k)$ measurement settings. The number of statistical repetitions is independent of $N$. Therefore, for $k=1\text{ and }2$ qubit subsystems, the scaling is $\mathcal O(N)$ and $\mathcal O(N^2)$ respectively~\cite{madhusudhana2023benchmarkingmultiqubitgates, madhusudhana2023benchmarkingmultiqubitgates2}.

\section*{ACKNOWLEDGMENT}
We acknowledge Professor Hartmut H{\"a}ffner for insightful discussions, valuable ideas, steady encouragement, and for providing access to the experimental setup used in the experimental part of this work.
This work was supported by the National Science Foundation (NSF)
Quantum Leap Challenge Institutes (QLCI) program through
Grant No. OMA-2016245, and by the U.S. Department of
Energy, Office of Science, National Quantum Information
Science Research Centers, Quantum Systems Accelerator.
\appendix

\section{CM and the PTM}\label{app:CMtoPTM}

We describe the mapping between the CM and the PTM in this section.

\begin{equation}\label{rho_toT}
    \rho^{\Phi} = \frac{1}{2}\sum_{\alpha \beta} T^{\Phi}_{\alpha\beta}\sigma_{\alpha}\otimes \sigma_{\beta}
\end{equation}

And
\begin{equation}
    T^{\Phi}_{\alpha\beta} = \frac{1}{2}\text{Tr}[\sigma_{\alpha}\otimes\sigma_{\beta}\rho^{\Phi}]
\end{equation}

The final state $\Phi(\rho)$ corresponding to an initial state $\rho$ can be computed from the CM using:
\begin{equation}
    \Phi(\rho) = \text{Tr}_2[\rho^{\Phi}\mathbbm{1}\otimes \rho^T]
\end{equation}
To see this, we begin with the definition of $\rho^{\Phi}$
\begin{equation}
    \begin{split}
        \rho^{\Phi} \mathbbm{1}\otimes \rho^T& = \sum_{ij}\Phi(\ket{i}\bra{j})\otimes \ket{i}\bra{j} \mathbbm{1}\otimes \rho\\
        & = \sum_{ij} \Phi(\ket{i}\bra{j})\otimes (\ket{i}\bra{j}\rho^T)
    \end{split}
\end{equation}
Taking a partial trace over the second subsystem:
\begin{equation}
    \begin{split}
        \text{Tr}_2[\rho^{\Phi} \mathbbm{1}\otimes \rho] & = \sum_{ij}\Phi(\ket{i}\bra{j}) \text{Tr}[\ket{i}\bra{j}\rho^T] \\
        & = \sum_{ij} \Phi(\ket{i}\bra{j})\rho_{ij} = \Phi(\sum_{ij}\ket{i}\bra{j}\rho^T_{ji})\\
        &=\Phi(\rho)
    \end{split}
\end{equation}

For a PTM, let $\vec{p}_{\rho}$ be the $4D$ vector representing the state in the pauli basis. The $4D$ vector of the final state is given by
\begin{equation}
    \vec{p}_{\Phi(\rho)} = T^{\Phi}\vec{p}_{\rho}
\end{equation}
To see this, let $\vec{p}_{\rho} = (\rho_0, \rho_x, \rho_y, \rho_z)$ 
\begin{equation}
    \rho = \rho_{\alpha}\sigma_{\alpha} \implies \Phi(\rho) = \rho_{\alpha}\Phi(\sigma_{\alpha})
\end{equation}
Therefore, 
\begin{equation}
\begin{split}
    \frac{1}{2}\text{Tr}[\sigma_{\beta}\Phi(\rho)] &= \sum_{\alpha}\frac{1}{2}\rho_{\alpha}\text{Tr}[\sigma_{\beta}\Phi(\sigma_{\alpha})] \\
    & = \sum_{\alpha} T^{\Phi}_{\beta\alpha}\rho_{\alpha} 
\end{split}  
\end{equation}
Note that the LHS is $\frac{1}{2}\text{Tr}[\sigma_{\beta}\Phi(\rho)] = (\sigma_{\beta}\Phi(\rho))_{\beta} $, the $\beta$ component of the corresponding $4D$ vector.  Thus, it follows that the $4D$ vector the final state is simply the matrix-vector product of $T^{\Phi}$ and the $4D$ vector of the initial state.

\section{Mitigating the measurement setting errors}

In Eq.~\ref{eq:y_2}, the operators $Y_i=T_{A_i} \vec{p}_{\ket{0} }
 \vec{p}^T_{\ket{1}} T_{B_i}$ may be inaccurate due to errors in the implementation of $A$ and $B$ which we refer to as \textit{measurement setting errors}. We follow the following procedure to mitigate these errors.

 \begin{itemize}
     \item[i.] We perform a process tomography \textit{without} the gate $U$ (i.e., $U=\mathbbm{1}$). The ideal PTM is $T^U=T^{\mathbbm{1}}=\mathbbm{1}$. 
     \item[ii.] Using the process tomography results, we obtain the experimental PTM $T_{\text{expt.}}^{(0)}$, using the procedure described in the main text, i.e., minimizing $f$ in Eq.~\ref{eq:y_3} under the PSD constraint. If there are no errors, we expect $T_{\text{expt.}}^{(0)}=\mathbbm{1}$
     \item[iii.] After a tomography with the gate $U$, we model the error as an operator acting before $U$. That is, we use the below equation instead of Eq.~\ref{eq:y_2}
     \begin{equation}\label{eq:y_4}
          y_i^{(0)} = \text{Tr}[Y_i T_{\rm expt.}^{(0)}T_{\rm expt.}],
     \end{equation}
     where $y_i^{(0)}$ denotes measurement outcome $i$, \textit{without} the gate $U$, and $Y_i$ the effective composite channel for this outcome.
     \item[iv.] Accordingly, we obtain $T_{\rm expt.}$ by minimizing
     \begin{equation}\label{eq:y_5}
         f(\tilde{T}) = \sum_i | \text{Tr}[Y_i T_{\rm expt.}^{(0)}\tilde{T}]-y_i|^2,
     \end{equation}
     under the PSD constraint, i.e., under the constraint that $\tilde{T}$ represents a completely positive map.
 \end{itemize}

\section{Numerical simulations}\label{app:numerical_sim}
To model the dynamics of the single-qubit tomography experiment, we employ a Monte-Carlo wave-function simulation, adding gate imperfections and measurement noise. No dissipation was considered during idling periods. Each trajectory begins in the pure state $\ket{0}$ and evolves over a sequence of discrete time steps. Whenever a control gate is scheduled, the state is updated by the corresponding ideal unitary operator. In addition, the selected gate at $t=3$, the rotation  $R_x(\theta)=e^{-i\theta X/2}$ with $\theta=\Omega t$ is perturbed by either a fixed miscalibration of the rotation angle to simulate coherent errors, or by a random fluctuation drawn from a distribution (Gaussian, of "top hat") to simulate non-Markovian errors.\\
To model the systematic over- or under- rotations discussed in Sec.\ref{sec:coherent}, the ideal rotation is replaced by a deterministic miscalibration 
$\theta \rightarrow \theta(1+\epsilon )$ with $\epsilon$ held constant through the entire simulation
This coherent error is identical for every trajectory and for every measurement shot, reproducing the effect of a incorrectly calibrated $\pi$-time or pulse area that persists across the full tomographic set.
To model the slow, shot-to-shot fluctuations described in Sec.\ref{sec:nonMarkov}, we instead allow the rotation angle to vary randomly between trajectories while remains constant within each single measurement sequence.
Measurements in the computational basis are implemented using a two-element POVM that incorporates readout errors, and the post-measurement state is projected onto the corresponding basis vector. After evolving 
$N$ trajectories, we construct the final density matrix as the ensemble average of the outer products of the final pure states and extract measurement statistics from the sampled outcomes.

\bibliography{references}

\end{document}